\documentclass[a4paper,aip,jcp,citeautoscript,reprint]{revtex4-2}

\usepackage[T1]{fontenc}
\usepackage[utf8]{inputenc}

\usepackage{booktabs}
\usepackage{threeparttable}
\usepackage[d]{esvect}
\usepackage{verbatim}
\usepackage[version=4]{mhchem} 
\usepackage{float}
\usepackage{pgfplots}
\usepackage[normalem]{ulem}

\usepackage{graphicx}
\usepackage{hyperref}
\hypersetup{
  colorlinks,
  citecolor=blue,
  linkcolor=blue,
  urlcolor=blue
}
  
\usepackage{siunitx}
\DeclareSIUnit\bar{bar}
\DeclareSIUnit\angstrom{\text {Å}}
\DeclareSIUnit\bohr{\text {b}}
\DeclareSIUnit\calorie{cal}

\usepackage{textgreek}
\usepackage{chemmacros}
\usepackage{amssymb}

\usepackage{float}

\begin{document}

\title{Extending Hamiltonian-Adaptive Resolution Simulation to Interfaces: An Updated LAMMPS Implementation and Application to Porous Solids}

\author{Hari Haran Sudhakar}
\affiliation{Sorbonne Universit\'e, CNRS, Physicochimie des \'Electrolytes et Nanosyst\`emes Interfaciaux, F-75005 Paris, France}

\author{Alessandra Serva}
\email{alessandra.serva@sorbonne-universite.fr}
\affiliation{Sorbonne Universit\'e, CNRS, Physicochimie des \'Electrolytes et Nanosyst\`emes Interfaciaux, F-75005 Paris, France}
\affiliation{R\'eseau sur le Stockage Electrochimique de l'Energie (RS2E), FR CNRS 3459, 80039 Amiens Cedex, France}

\author{Rocio Semino}
\email{rocio.semino@sorbonne-universite.fr}
\affiliation{Sorbonne Universit\'e, CNRS, Physicochimie des \'Electrolytes et Nanosyst\`emes Interfaciaux, F-75005 Paris, France}

\date{\today}

\begin{abstract}
Many natural phenomena involve processes that happen simultaneously at different characteristic length- and timescales. Typically, the region where the process of interest happens is affected by fluctuations in its surroundings. Modeling these systems requires an effective combination of simulation resolutions. The Hamiltonian-Adaptive Resolution Simulation (H-AdResS) method allows to model dual-resolution systems in length- and time-scales compatible with molecular diffusion, by combining atomistic and particle-based coarse graining models in the same simulation box. In this work, a new implementation of H-AdResS is provided in LAMMPS 2023. New features extend the usage to more diverse interaction potentials and simplify the preparation of input files via dedicated lammps input commands, while keeping the efficiency gain of the basis method. The implementation is benchmarked by reproducing water properties from a reference atomistic simulation. Importantly, the new implementation includes changes in compensation routines allowing to simulate systems with fluctuating density. As an example, the method in its new implementation is applied to modeling a porous metal-organic framework and its gas adsorption structure and transport properties. We demonstrate that structural and dynamic properties in the atomistic region of the dual-resolution scheme are unaffected and remain those of the fully atomistic system, while increasing simulation efficiency. This paves the way for using H-AdResS to simulate complex interfaces across applications in energy storage, electrocatalysis, and membrane technologies. \\

\textbf{Keywords:} H-AdResS; multiscale simulation; metal-organic frameworks; lammps; molecular dynamics

\end{abstract}

\maketitle

\section{Introduction}

Computer simulation techniques have been instrumental in the development of physical, chemical and biological sciences in the past half century as they provide access to molecular-level degrees of freedom that are often difficult to sample via direct experimental measurements. Systems of interest include chemical reactions, protein folding and interfacial or confined fluid structure and reactivity, among others. These varied examples can all be described in a more abstract way as composed by a region in which the crucial phenomenon under investigation occurs, and an environment that surrounds it. The environment plays an important role by modulating the fluctuations that will eventually lead to the phenomenon of interest. The physico-chemical behaviour of this kind of systems is often dictated by a delicate interplay between processes occurring in a wide range of characteristic time- and length-scales. Modeling them involves taking into account all these time- and length-scales simultaneously, which, unfortunately, cannot be done via expensive, high resolution quantum mechanical models.

As a result, the concept of multiscale simulation emerged. Within these approaches, two (or more\cite{Chung2015}) simulation resolutions are combined and linked to capture multiple scales of the system of interest at the same time.\cite{Peter2010} Early developments in the field included merging quantum mechanical models with classical atomistic models (AA), in hybrid schemes called QM/MM methods.\cite{Warshel1976} However, the timescales at which these methods typically operate are not compatible with molecular diffusion. The Adaptive Resolution Simulation method (AdResS) provides an elegant solution to this conundrum, by moving on to lower resolutions and allowing free diffusion of molecules throughout the whole system.\cite{Praprotnik2008_2,Praprotnik2020} 

\begin{figure}[h]
    \centering
    \includegraphics[width=0.5\textwidth]{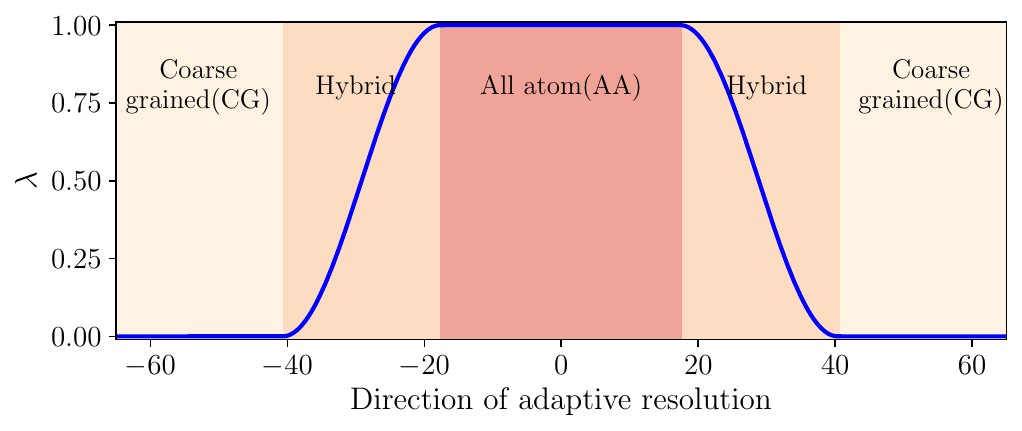}
    \caption{Schematic representation of the Hamiltonian-Adaptive Resolution Simulation setup. Three different regions, namely (i) atomistic (AA) (ii) coarse grained (CG) and (iii) hybrid (connecting the first two), are defined. Molecules are allowed to move from one region to another and the instantaneous resolution of the particle is given by the transition function according to equation (\ref{eq:transition}) and illustrated in blue.}
    \label{fig:hadress-scheme}
\end{figure}

In the AdResS scheme,\cite{Praprotnik2005} the portion of the system that contains the crucial chemistry is treated at the AA resolution, while the environment is typically treated via a lower resolution particle-based coarse grained (CG) model.\cite{Reith2003,Souza2021} A hybrid region lies in-between the high- and low-resolution portions of the system, in which particles possess a dual-resolution identity. A sketch of this set-up is shown in Figure \ref{fig:hadress-scheme}. Forces over molecules that belong to the hybrid region are linearly interpolated from those corresponding to the single-resolution portions of the system weighted by the instantaneous relative distance of the molecule to these regions. Another variant of this method proposes direct coupling between AA and CG regions, without any hybrid region in-between.\cite{Krekeler2018,Thaler2020}

An alternative formulation is the Hamiltonian one, H-AdResS,\cite{Potestio2013} in which the interpolated quantity is the non-bonded contribution of the potential energy of the molecules lying in the hybrid region, instead of the forces. Both schemes have been implemented in the molecular dynamics (MD) code ESPResSo++\cite{Halverson2013} (see \url{http://www.espresso-pp.de/Documentation/tutorial.AdResS.html?highlight=adress}), AdResS in GROMACS\cite{Berendsen1995} (see \url{https://manual.gromacs.org/documentation/5.1.4/doxygen/html-full/adress_8h.xhtml}) and H-AdResS in LAMMPS\cite{Thompson2022} 2016.\cite{Heidari2016} Among those open source molecular simulation packages, LAMMPS has been developed with a material's modeling mindset and has been widely used by the community. Since 2016, LAMMPS has undergone major changes, including an upgrade to modern C++ 11, changes in the way read and write utilities work and the introduction of new fix/compute styles.
For this reason, the 2016 H-AdResS implementation cannot be directly transposed into the newer LAMMPS versions, and using it becomes impractical.

Even though AdResS and H-AdResS schemes were originally developed without any particular system specificity in mind, most applications to date are focused on bulk liquids, liquid mixtures and solutions.\cite{Praprotnik2006,Praprotnik2008,Lambeth2010,Potestio2013,Mukherji2013,Fogarty2015,Fogarty2016,Zavadlav2017,ShadrackJabes2018,CortesHuerto2021,Oestereich2024} No bulk gas nor gas/solid systems were ever explored via these schemes to the best of our knowledge, and we could only find one solid state application\cite{Heidari2018} to date, despite the fact that possible applications abound, including the broad field of gas adsorption in porous materials.

The 2025 Nobel Prize in Chemistry was awarded for the discovery of metal-organic frameworks (MOFs), porous materials that combine metal ions or oxoclusters with organic polydentate molecular ions,\cite{Furukawa2013,Zhou2014} and that have remarkable gas storage and separation properties.\cite{Li2018,Li2022} Current gas adsorption simulation efforts rely mostly on AA models,\cite{Daglar2024} although some recent works have explored CG models for this purpose.\cite{Alvares2024,Mohamed2024,Alvares2025,Alvares2026} Applying AdResS or H-AdResS schemes to these systems would allow to keep AA resolution, which contains molecular detail, in a portion of the system, while treating larger systems overall. This would open the path to explore the impact of variables such as concentration of defects in the MOF or MOF-based composite composition on adsorption properties. 

In this work, we first provide an implementation of the H-AdResS method in a recent version of LAMMPS (version 2 Aug 2023 - Update 2). This implementation, made openly available in \url{https://github.com/hariharansudhakar/lammps-hadress}, includes optimized AA region pair styles, new pair styles to treat the CG region and new i/o commands that make the method easier to use. We test our implementation by comparing structure and dynamic properties of bulk water modeled in a fully AA fashion with those computed for the AA region of a H-AdResS simulation, as well as with the 2016 H-AdResS implementation. Once validated, we apply the H-AdResS method to modeling a bulk MOF and MOF/\ce{CO2} interfaces. This is, to the best of our knowledge, the first time this method is applied to porous solids and to porous solid/gas interfaces. We demonstrate that structural and dynamic properties in the AA region are not perturbed by the H-AdResS scheme, thus validating the H-AdResS scheme to studying porous solids and porous solids/gas interfaces.

The article is structured as follows. Section II describes the method. Section III describes implementation of H-AdResS in LAMMPS 2023 (version 2 Aug 2023 - Update 2). Section IV covers all simulation setups, while Section V contains the results. Conclusions are summarized in Section VI.

\section{Method}
In the H-AdResS scheme\cite{Potestio2013}, every particle in the simulation box interacts with its neighbours via either CG or AA potentials or a combination of both. These potentials can have any functional form and interaction order (two-body, three-body, etc.) as long as the extension is short-ranged. The interactions among the particles in the simulation box is given by the global Hamiltonian:
\begin{equation} \label{eq:hamiltonian}
H  = \kappa + U_{bonded} + \{\lambda_a . U^{AA}_a + (1-\lambda_a).U^{CG}_a \}
\end{equation} 

The term $\kappa$ denotes the kinetic energy, $U_{bonded}$ denotes the potential energy contributions from the bonded interactions. $U^{AA}$ and $U^{CG}$ denote the non-bonded potential energy of the molecule when it is in AA or CG region, respectively. For the most common case of pairwise interactions, from \eqref{eq:hamiltonian} the non-bonded interactions for a molecule $a$, are the weighted sum of $U^{AA}$ and $U^{CG}$ given as,
\begin{equation}\label{eq:atomistic}
U_{a}^{AA} \equiv \frac{1}{2} \sum_{\substack{b,\, b \neq a}}^{N} \sum_{ij} U^{AA}\!\left(|x_{ai} - x_{bj}|\right)
\end{equation}

\begin{equation}\label{eq:coarse-grainned}
U_{a}^{CG} \equiv \frac{1}{2} \sum_{\substack{b,\, b \neq a}}^{N} U^{CG}\!\left(|X_{a} - X_{b}|\right)
\end{equation}

where $x_{ai}$ and $x_{bj}$ denote the position of atoms $i$ and $j$ that belong to molecules $a$ and $b$ respectively. Both \eqref{eq:atomistic} and \eqref{eq:coarse-grainned} are calculated and scaled according to their $\lambda$ value for interactions within the hybrid region and cross resolution interactions. $\lambda$ is a function of the center of mass ($X$) of the molecule $a$. For any atom $i$ in molecule $a$, $\lambda_i = \lambda(X_a)$. This denotes the resolution of the particle in the simulation box, which can take any values between 0 (fully CG) and 1 (fully AA). Since the atomistic degrees of freedom are preserved along the simulation, bonded terms are treated at the AA level for CG molecules as well, as established in the H-AdResS Hamiltonian (equation \ref{eq:hamiltonian}). Thus, no new routines are needed to compute the bonded interactions in LAMMPS. The force acting on an atom $i$ derived from the H-AdResS Hamiltonian is,

\small
\begin{equation}\label{eq:force}
\begin{split}
F_{ai} &= F_{ai}^{bonded} + \sum_{\substack{b,\,b \neq a}} 
\left\{ \frac{\lambda_a + \lambda_b}{2} F_{ai|b}^{AA} + \left(1 - \frac{\lambda_a+ \lambda_b}{2}\right) 
F_{ai|b}^{CG} \right\} \\
&\quad - \left[U_{a}^{AA} - U_{a}^{CG}\right] \nabla_{ai} \lambda_a.
\end{split}
\end{equation}
\normalsize

The first part of equation \eqref{eq:force} represents the force due to the bonded interactions. The second term represents the force experienced by atom $i$ of molecule $a$ due to its interaction with molecule $b$. The third term is the drift force term, that arises from the discontinuity in resolution. AA potentials are parameterized in such a way that they represent structure and chemical specificity whereas CG potentials are parameterized to reproduce atomistic properties, such as forces or structure at a particular temperature and pressure, considering a collective degrees of freedom.\cite{Rzepiela2011, Faller2004, Agrawal2016} As a result, they are usually not transferable to other system compositions or thermodynamic conditions. Thus, AA and CG potentials are fundamentally different. Therefore, at a given thermodynamic state (T,V), the models can yield different pressure in the AA and CG resolutions,\cite{Carbone2008} resulting in non-homogeneous densities. This leads to a selective diffusion of particles to the region where the Helmholtz free energy is lower in the canonical ensemble. To overcome this thermodynamic imbalance, compensation functions are introduced, which are function of the position of the center of mass of the molecules. 
\begin{equation}\label{eq:modified-hamiltonian}
H' = H - \sum_{a=1}^{N} \Delta H(\lambda(X_{a})).
\end{equation}

This term removes the average drift force acting on individual molecules in the hybrid region. To compensate the drift force $F^{drift}_a$, it should obey the relation:
\begin{equation}\label{eq:drift-force}
    \frac{d \Delta H(\lambda)}{d\lambda}\bigg|_{\lambda = \lambda_a} = <[U_a^{AA} - U_a^{CG}]>\bigg|_{X_a}
\end{equation}

By construction, <$F^{drift}_a$> will be equal to zero in the hybrid region, balancing hydrostatic pressure gradient across the regions. This quantity can be pre-computed via thermodynamic integration\cite{Kirkwood1935} techniques. But the most accurate and efficient way is to compute and update the compensation terms dynamically, as an iterative process.\cite{Heidari2016} In cases where density should be uniform besides or instead of pressure, the chemical potential gradient in the hybrid region has to compensated. Similar to the pressure compensation routine, density compensation forces can also be computed via an iterative scheme\cite{Fritsch2012}:
\begin{equation}\label{eq:density-comp}
    F_{n+1} = F_n + \frac{c.\nabla \rho_n(x)}{\rho^*}
\end{equation}
 
 The $\nabla\rho$ term in the expression ensures the convergence to uniform density and c is a prefactor with energy units.  Applying the sum of forces from pressure and density compensation routines ensures the chemical potential gradient is counterbalanced. This strategy ensures a flat density profile in the hybrid region while maintaining same density both in AA and CG regions (see Figure \ref{fig:dens-water}). 

\section{Implementation}

\begin{figure}[htbp]
    \centering
    % \textwidth scales it to fit the width of your text column
    \includegraphics[width=0.5\textwidth]{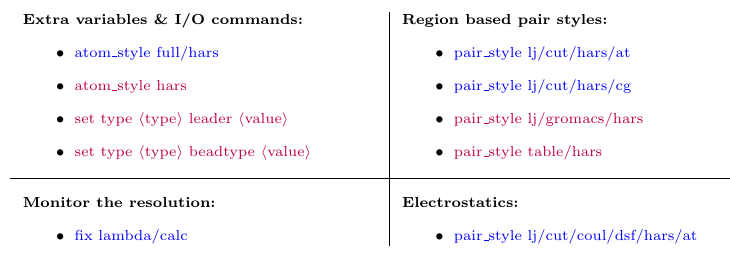}
    \caption{Commands and functions highlighted in blue were present in the 2016 implementation and migrated into our implementation. The commands highlighted in red are the new functions introduced in our implementation. }
    \label{fig:my_pdf_table}
\end{figure}

In this section, we begin with introducing H-AdResS features that were adapted from the earlier 2016 implementation,\cite{Heidari2016} followed by the extensions and improvements made in our current implementation. A schematic summary is provided in Figure \ref{fig:my_pdf_table}.

To model a system in dual resolution, routines specific to H-AdResS are needed, along with modifications to the traditional MD algorithms, so that they can accommodate the changes in resolution schemes. First, the atom style \textit{hars} \texttt{(class AtomVecHars)} was introduced, as in the 2016 implementation\cite{Heidari2016}. This style was adapted from the existing style \textit{full}\cite{fullstyle} \texttt{(class AtomVecFull)}. It contains the new per-atom variables required for the scheme along with pre-existing per-atom quantities such as positions, charges and velocities. For every atom present in the simulation, the following variables were introduced: (i) $X$: Center of mass of the molecule to which the atom belongs to; (ii) \textit{Leader atom flag}: Every molecule contains one leader atom, that acts as a point of contact and contains all the information about the molecule, this flag allows to know which atom is the leader atom, and which are not; (iii) $CG_{type}$: It takes values from 1 to n, where n is the total number of types of CG beads present in the CG region; (iv) $\lambda $: This variable stores the resolution of the particle [0-1], as shown in Figure \ref{fig:hadress-scheme}. It is defined by the transition function (\ref{eq:transition}) computed using the position of the center of mass of the molecules;
(v) $\nabla \lambda$: Gradient of the transition function. With this atom style, all the atoms present in the simulation box will contain the information of which region they are in, which type of CG bead and which particular group of atoms they belong to, during the course of the simulation. 

Once the \texttt{hars} style defined, we define the interactions between the atoms, which are governed by different pair styles for different regions. For the AA region, all the interactions are pairwise and short-ranged, and they are governed by the \textit{lj/cut/coul/dsf/hars/at} style \texttt{(class PairLJCutCoulDSFHARSAT)} derived from the \textit{lj/cut/coul/dsf} style\cite{DSFpotential} native to LAMMPS. This pair style computes Lennard Jones (LJ) interactions and Damped Shifted Force (DSF) \cite{Fennell2006} model for Coulombic energies and forces for atoms in the AA region as well as for the atoms in the hybrid region. This pair style definition requires a damping parameter, a global cutoff for LJ and Coulombic interactions and a (restart) flag that can be switched on in the case of restarting a H-AdResS simulation, so that it can read the compensation energy files written from the previous run to maintain the continuity. For the CG region, interactions are governed by the style \textit{lj/cut/hars/cg} \texttt{(class PairLJCutHARSCG)}.
   
During the simulation, the positions of the particles evolve and so does the instantaneous resolution of the molecules or fragments. Thus, this change in $\lambda$ is monitored and updated via a fix named \textit{lambda/calc} \texttt{(class class FixLambdaHCalc)}. This fix computes the center of mass of each molecule and then updates their $\lambda$ and $\nabla \lambda$. Also, calling this fix in the input triggers the initialization and memory allocation for all the variables used by the \texttt{(class FixLambdaHCalc)} routines during the simulation and then broadcasts these variables and flags to the pair style routines. The $\lambda$ values are continuous and calculated using the following position based piece-wise function:
\begin{equation}\label{eq:transition}
\lambda(x) = \begin{cases} 
1 & |x| \leq l_{AA}/2 \\
\cos^2 \left( \frac{\pi (x - l_{AA}/2)}{2 l_{HY}} \right) & l_{AA}/{2} < |x| \leq l_{AA}/{2} + l_{HY} \\
0 & |x| > l_{AA}/2 + l_{HY}.
\end{cases}
\end{equation}

 In general, for the AA region $\lambda $ takes the value of 1 and for the CG region, it takes the value of 0. In the hybrid region, it takes values between 0 and 1. Mathematically, the values depend on the size of the atomistic ($l_{AA}$) and hybrid ($l_{HY}$) regions. The size of these regions are user-defined and initialized through the LAMMPS input script.
 
 We extended the implementation by introducing new features and functions. Two new pair styles \textit{lj/gromacs/hars} \texttt{(class PairLJGromacsHARS)} and \textit{table/hars} \texttt{(class PairTableHARS)} were developed\cite{lammpsbook}. With these new pair styles, the implementation is more flexible, allowing to use different forms of force fields, either an analytical potential such as simple LJ,\cite{ljcut} tabular potentials developed via any CG-force field-fitting method such as Iterative Boltzmann Inversion (IBI)\cite{Reith2003} and Multiscale Coarse-Graining (MS-CG)\cite{Izvekov2005} or general purpose CG force fields such as MARTINI,\cite{Marrink2007,Souza2021} to govern the interactions in the CG region. Since our implementation can be used with tabular potentials, any analytical potential can be used for the CG region, as long as it can be transformed into a table form. As mentioned above, the interactions between the atoms in the hybrid region and the case where one atom is in the hybrid region and the others in the AA or CG regions, are computed by interpolation of the potentials corresponding to the AA and CG regions. These contributions are calculated by the respective pair styles used for the region and are scaled by the transition function.

To make the method more accessible, we created two i/o commands under the \texttt{class Set}. The usage is similar to the other \textit{set} commands. The syntax is as follows: \\ 
\textit{set type <atom type> leader <flag>} \\ 
\textit{set type <atom type> beadtype <beadtype>} \\

With these new commands, the leader atom flags and CG bead type for each atom can be easily assigned through the LAMMPS input script, removing the need of writing an external script to prepare the input LAMMPS data file. This allows us to explore different CG mappings and change their leader atom flags easily, using few lines of simple commands. This reduces time and effort when trying multiple CG mappings for the same system. 

Pressure and density compensation routines were mostly implemented as in the 2016 version (see SI). Usually the $\delta \lambda$ value is set to be very small, so that the number of bins are large enough and therefore the transition from one region to another is smooth and continuous. For systems that do not have uniform density across the box, such as gases and ordered porous solids like MOFs, this method of counting the particles in a bin can result in discontinuity and unphysical energies/forces, when it encounters a bin without any particles. In our implementation, the compensation routines were modified to handle this inconsistency. The algorithm is re-written in such a way that every time an empty bin is encountered, it will be assigned an energy interpolated from its adjacent bins. By doing so, when an atom enters a bin which was empty at previous step of integration, a physically meaningful compensation energy will be applied on it, unlike in the previous case. This extends the usability of H-AdResS to a wider variety of systems with non-homogeneous density. 

While simulating larger molecules, the CG representations of the molecule are generally made up of more than one CG bead. In such scenario, for a same molecule, the fragments may lie in different resolution regions. These molecular fragments that lie at the boundary of two regions undergo thermal fluctuations which could lead to a sudden re-introduction/removal of AA degrees of freedom, resulting in unphysical energies/forces in the hybrid region (especially when molecules lie between hybrid and CG regions). This may even lead to crashing simulations. In order to overcome this issue, a force and energy capping mechanism is introduced in our implementation in the \textit{compute} routines of the pair styles, to dissipate the high energies/forces experienced in the boundaries of hybrid regions.

The native LAMMPS codes are largely unaffected. Very few classes such as \texttt{Class Atom}, \texttt{Class Molecule} and \texttt{Class Set} were modified to integrate the H-AdResS routines. All the other methods such as atom style, pair styles and fixes are available in the folder \texttt{HADRESS} in the \textit{src} directory. All the H-AdResS parameters such as H-AdResS switch on/off flags, compensation routines on/off flags, compensation update frequency, pressure and density compensation parameters  are given as input through the LAMMPS script via the fix \textit{lambda/calc} command. The compensation energies calculated by the AA and CG pair styles are updated at a time frequency given as an input, and these quantities are written in a file, which allows to restart a H-AdResS simulation from any given time. Flags can be switched on via pair style and fix commands to read these files.

\section{Simulation details}

In order to test our implementation, we began by performing a simulation of bulk liquid water. We then moved to the simulation of a MOF, the Zeolitic Imidazolate Framework ZIF-8,\cite{Park2006} both empty and loaded with \ce{CO2}, as described below. Simulation settings including cut-offs, thermostats, damping constants and the nature of the H-AdResS parameters as well as sensible choice of those values are reported in the SI.

\subsection{Liquid water}
A rectangular box of $200$ \AA \ x $40$ \AA \ x $40$ \AA \ was filled with 10240 water molecules using the \textit{create\textunderscore atoms} function in LAMMPS. The SPC/E water model\cite{Berendsen1987} was used for AA resolution, while the Weeks-Chandler-Andersen (WCA) potential\cite{Weeks1971} was employed for the CG one. The initial configuration generated by LAMMPS was equilibrated in the NPT ensemble at 300 K and 1 atm for 500 ps. Then, a 2 ns production run was performed in the NVT ensemble at 300 K. For H-AdResS simulations, an equilibrated configuration generated from an AA simulation was used as initial configuration. The size of AA and hybrid regions were set to be 60 \AA \ and 25 \AA, respectively. After a 50 ps-long run, either one or both compensations were activated. The equilibration was performed for 500 ps. Then, a 2 ns production run was performed. For pressure compensation, $\nabla\lambda$ was set to be 0.02 and for density compensation, bin width was set to 1.5 \AA. A Langevin thermostat with 0.1 ps damping constant was used for the whole box.

%%%%%%%%%%%%%%%%%%%%%%%%%%%%%

\subsection{ZIF-8}\label{subsec:solid-mof} 

ZIF-8\cite{Park2006} is made up of Zn$^{2+}$ as metal centers and 2-methylimidazolate as organic linkers. A single bulk MOF crystal of $ 135.68$ \AA \ x $33.92$ \AA \ x $33.92$ \AA \ (8x2x2 supercell) was used. The interactions among ZIF-8 atoms were modeled using ZIF-FF\cite{Weng2019} and a MARTINI-based force field\cite{Alvares2023} for the AA and CG resolutions respectively. The initial configuration was generated using a python script and equilibrated in the NPT ensemble at 300 K and 1 atm. Then, a 2 ns production run was performed in the NVT ensemble at 300 K. For H-AdResS simulations, the size of AA and hybrid regions were set to be 40 \AA \ and 25 \AA \ respectively. After a 50 ps-long run, compensations were activated. The equilibration was performed for 500 ps. Then, a 2 ns-long production run was performed. For pressure compensation, $\nabla\lambda$ was set to 0.25 and for density compensation, bin width was set to 2.0 \AA. Two Langevin thermostats, one for AA+hybrid regions and another one for the CG region, with 0.1 ps damping constant, were used. 
%%%%%%%%%%%%%%%%%%%%%%%%%%%%%

\subsection{ZIF-8/\ce{CO2}}

We further simulated ZIF-8 loaded with \ce{CO2}. The MOF was modeled in the same way as discussed in the section \ref{subsec:solid-mof}. For AA resolution, \ce{CO2} was modelled using EPM-2 model\cite{ZHU2009} and for CG resolution, a MARTINI-based force field\cite{Alvares2025} was used. The gas was loaded into the MOF using the Grand Canonical Monte Carlo (GCMC) method. From experimental studies, it has been observed that, at 300 K and 1 atm, ZIF-8 can take up 0.48 to 1.0 mmol/g of \ce{CO2}\cite{NeumannBarrosFerreira2024, Klomkliang2025, Wu2014, Zhang2014, Pusch2012}, which corresponds to 0.8 to 1.8 molecules of \ce{CO2} per unit cell of ZIF-8. To avoid any unphysical diffusion of gas between regions in H-AdResS simulations, full AA and full CG simulations were setup. \ce{CO2} loading reproduced by both the force fields were measured with GCMC+MD simulations. Both AA and CG force fields reproduce an identical loading of 1.5-2 \ce{CO2} molecules per unit cell when exposed to a pressure of 1 atm from the virtual gas reservoir (Figure \ref{fig:loading}).
 
\begin{figure}[h]
    \centering
    \includegraphics[width=9cm]{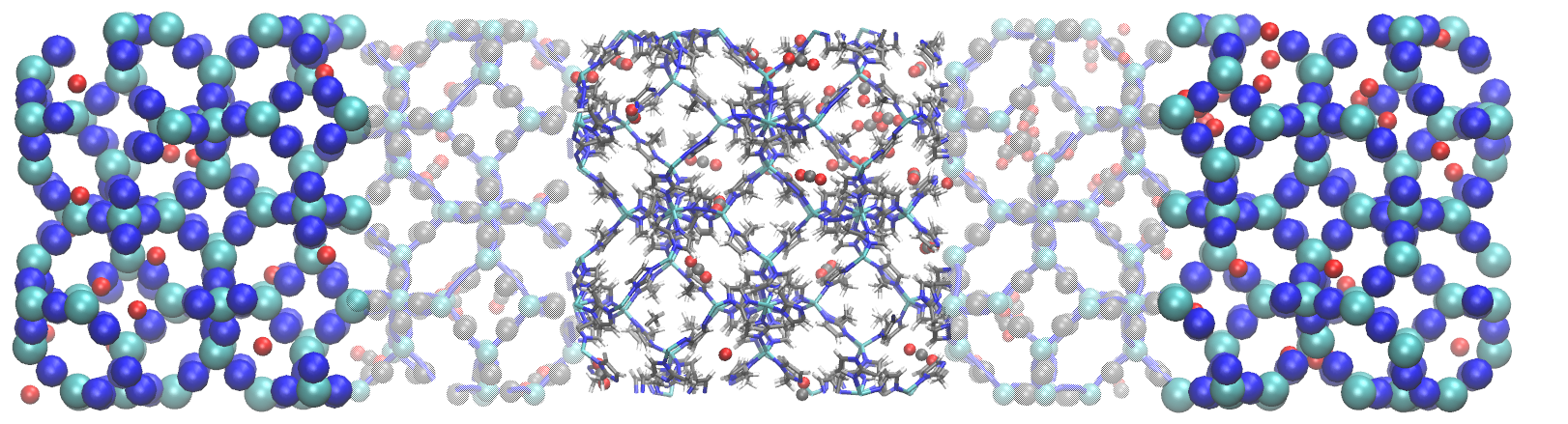}
    \caption{ZIF-8/\ce{CO2} system modelled in H-AdResS approach. In the CG region, \ce{CO2} molecules are shown in red, 2-methylimidazolate ligands of ZIF-8 in blue and Zn metal centers in cyan. Hybrid region is shown as translucent denoting the change in resolution in the region. In AA region, carbon atoms are shown in grey, hydrogen in white, nitrogen in cyan and oxygen of \ce{CO2} gas in red. The \ce{CO2} gas can freely diffuse and exchange between the regions. }
    \label{fig:zif-8-co2}
\end{figure}

Once the gas was loaded, the resulting configuration was equilibrated for 2 ns. A representative picture of an initial configuration for these simulations is shown in Figure~\ref{fig:zif-8-co2}. Langevin thermostat with 4 ps damping constant was used. Five independent 10 ns-long production runs were performed. Both for fully atomistic and H-AdResS cases, the results we report are an average of five independent MD runs. For H-AdResS simulations, the size of atomistic and hybrid regions were set to 40 \AA \ and 25 \AA \ respectively. After 50 ps run, compensations were activated. The equilibration was performed for 1 ns. Then, a 10 ns production run was performed. For pressure compensation, $\nabla\lambda$ was set to be 0.05 and for density compensation, bin width was set to be 1.5 \AA. Two Langevin thermostats, one for AA+hybrid regions and another one for CG region, with 1 ps damping constant, were used.

\section{Results and discussion}

\subsection{Validation}

\begin{figure}[h]
    \centering
    \includegraphics[width=0.5\textwidth]{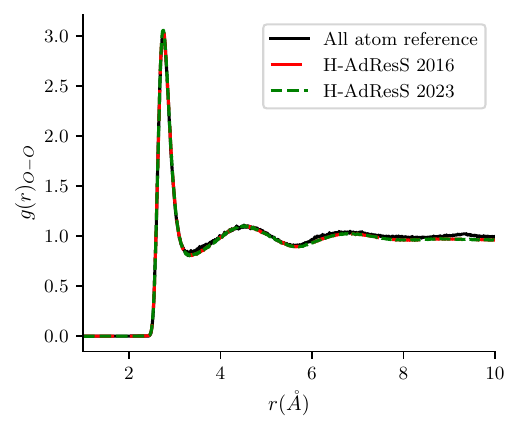}
    \caption{Radial distribution function, g(r), of the O--O pair in bulk liquid water computed from a full AA simulation (black solid line) used as a reference against the g(r) computed in the AA region of a H-AdResS simulation performed using the 2016 implementation (red long dashes) and our 2023 implementation (green dashes). H-AdResS simulations are performed with both the compensations active. }
    \label{fig:rdf-water}
\end{figure}

To evaluate the accuracy of our implementation of the H-AdResS scheme in LAMMPS 2023, we used liquid water as benchmark system, as previously done in the field\cite{Zavadlav2016,Zavadlav2017,Wang2018}. We computed radial distribution functions (RDFs) and density profiles along the x-direction (i.e the direction where resolution changes). RDFs from three different scenarios including, full AA, H-AdResS using the 2016 implementation \cite{Heidari2016} and H-AdResS simulation using our new implementation were compared and reported in Figure \ref{fig:rdf-water}. From the excellent agreement, it can be confirmed that our new implementation accurately reproduces structure results in the AA region of H-AdResS scheme.

\subsection{Application}

\begin{figure}[h!]
    \centering
    \includegraphics[width=0.5\textwidth]{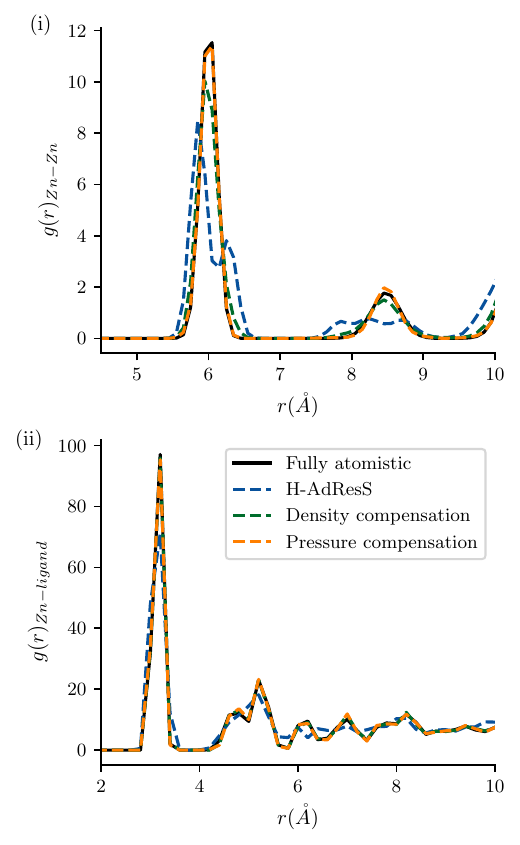}
    \caption{Radial distribution functions, g(r), obtained from the AA region of H-AdResS simulations with no compensations active (blue dashed line), with pressure compensation active (orange dashed line) and density compensation active (green dashed line) and the reference AA simulation (solid black line) of the bulk ZIF-8. g(r) between Zn--Zn and Zn-ligand pairs are reported in the top and bottom panels respectively.}
    \label{fig:rdf-bulk-mof}
\end{figure}

Once our H-AdResS implementation validated, we simulated bulk ZIF-8. To measure the impact of compensation functions on the structure, we performed four different simulations including full AA, H-AdResS with no compensations active and H-AdResS with either the pressure or the density compensations active. It can be seen that the sharp peaks corresponding to the crystalline nature of ZIF-8 are well reproduced in the presence of pressure compensation, compared to density compensation (see Figure \ref{fig:rdf-bulk-mof}). Due to the ordered and porous nature of ZIF-8, number density across the box of ZIF-8 will have alternating peaks and valleys throughout the simulation box. The role of density compensation is to reproduce a uniform density profile, which is a property that is inherent of bulk fluids. Thus,density compensation should not be needed to reproduce the structure of bulk ZIF-8.

\begin{figure}[h!]
    \centering
    \includegraphics{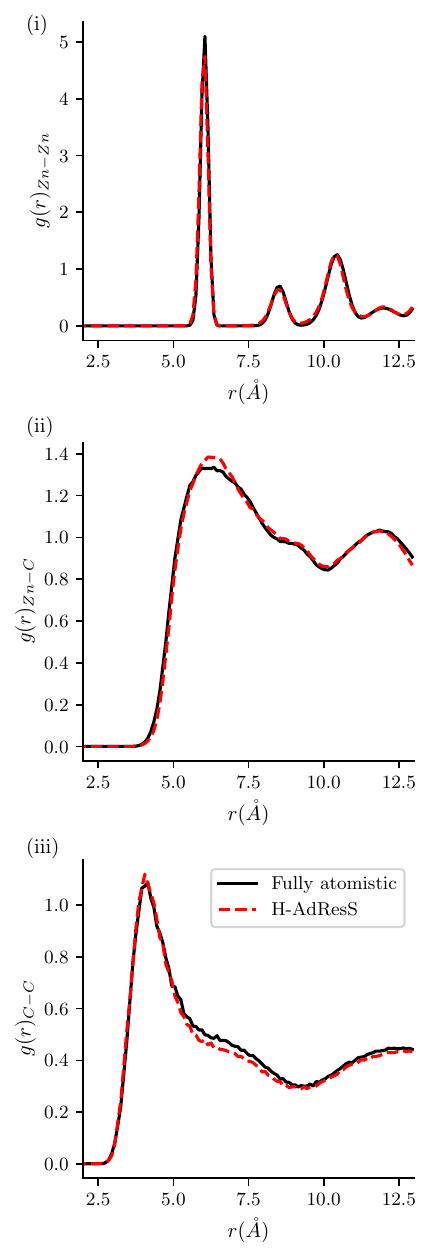}
    \caption{Radial distribution functions, g(r), of (i) Zn--Zn (ii) Zn--C (iii) C--C interactions, obtained from the AA region of H-AdResS simulations (red long dashes) and the reference AA simulations (solid black lines) of the ZIF-8/CO$_2$ system.}
    \label{fig:rdf}
\end{figure}

We then applied H-AdResS to study the ZIF-8/\ce{CO2} interface. With its exceptional structure tunability and proven potential to capture \ce{CO2} gas, ZIF-8 is a good material for membranes with applications ranging from industrial gas separation processes to climate change mitigation\cite{Abraha2021, Gong2017, Jiang2023, Vendite2022}. To assess whether the structural organization of the system is well-reproduced in the H-AdResS simulation, we started our analysis by computing RDFs between three different representative atom pairs for ZIF-8/\ce{CO2}: Zn--Zn atoms of ZIF-8, C--C atoms of \ce{CO2} and Zn--C atoms, both within the AA region of a H-AdResS simulation and in a fully AA simulation.
Figure \ref{fig:rdf} shows that both the first sharp peak and the subsequent peaks at longer distances are well reproduced in the Zn--Zn RDF, indicating that the crystalline nature of ZIF-8 is well preserved in the H-AdResS scheme. From the excellent overlap of the Zn--C and C--C RDFs, it is evident that the structural behavior of \ce{CO2} gas around the metal ions and the interactions between the gas molecules are also correctly reproduced in the AA region of the H-AdResS simulations. \\

\begin{figure}[t]
    \centering
    \includegraphics{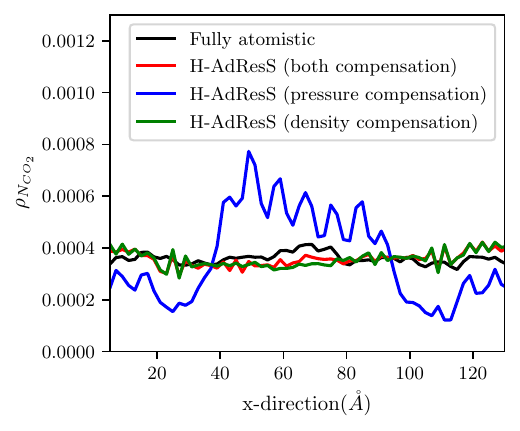}
    \caption{Density profile of \ce{CO2} gas in different simulation settings. Blue curve represents H-AdResS scheme with only pressure compensation active, green curve represents H-AdResS scheme with only density compensation active, red curve represents both compensations active. The black curve is obtained from the reference AA simulation. All these curves are plotted against the direction of the change in resolution (x-direction).}
    \label{fig:density-profile}
\end{figure}

As mentioned before, having a constant density across the different regions is key for bulk fluids under equilibrium conditions. We studied the density profile of \ce{CO2} in the x-direction, i.e., the direction of the change in resolution. The density profile is computed by dividing the simulation box into small bins and counting the total number of \ce{CO2} molecules present in each bin at the given time. The measured bin count is averaged over all the frames and divided by the volume of each bin. This results in the average number density profile of \ce{CO2} gas, reported in Figure \ref{fig:density-profile}. The bin size used to compute this profile is 2 \AA. The plot shows the importance of compensating both the drift force and density imbalances. Even without one of the compensations, the system evolves into a state where different regions reproduce different average densities. By applying both pressure and density compensation routines, uniform average density across the box is enforced, including in the hybrid regions and matching the one of the fully AA simulation.

Beyond the structural properties, to ensure that the dynamics in the AA region is also reproduced as compared to the reference AA simulation, we computed the mean residence time ($\tau_R$). It is defined as the average time spent by the gas molecule inside a pore. This is measured by the survival probability, a continuous time correlation function S(t) \cite{Snchez2022}, given by,

\begin{equation}\label{eq:time-correlation}
    S(t) = \frac{\left <h(0)H(t)\right>}{\left <h(0)h(0)\right>}
\end{equation}

with a tolerance time,\cite{Impey1983} t*, of 2 ps. For the reference AA simulation, $\tau_R$ of \ce{CO2} is  47 $\pm$ 4 ps, while a value of 47 $\pm$ 7 ps is obtained for the AA region within H-AdResS with both compensation routines active. These values are in very good agreement with those reported in the literature (44 ps\cite{Yang2013}) at the given temperature. It should be noted that with only pressure compensation the mean residence time is 97.29 ps while it is of 22.01 ps with only density compensation. This shows that both compensations are needed for this system. From the density profile (Figure \ref{fig:density-profile} - blue curve), it can be observed that the density of gas in the AA region is higher in the former case. This crowding of gas in the AA region created resistance for the gas molecules to exit the pores, thus explaining the higher $\tau_R$ found for systems in which only the pressure compensation is applied. In the case in which only density compensation is activated, though we apply thermodynamic forces, the gas tends to enter/exit the pores faster due to the unbalanced drift forces.

\begin{figure}[b]
    \centering
    \includegraphics[width=0.5\textwidth]{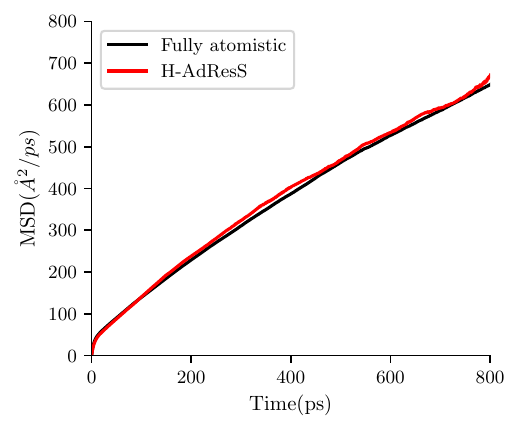}
    \caption{Mean squared displacement (MSD) of \ce{CO2} in the AA region of the ZIF-8/\ce{CO2} system computed as a function of time at 300 K.}
    \label{fig:msd}
\end{figure}

% \begin{figure}
%    \centering
%    \input{figs/rtd.pgf}
%    \caption{Survival probability of \ce{CO2} gas inside a ZIF-8 pore, computed using an continuous function S(t), plotted against simulation time.}
%    \label{fig:rtd}
%\end{figure}

Subsequently, we measured the \ce{CO2} diffusion coefficient as a second probe to assess the accuracy in reproducing dynamic properties from the benchmark AA system. We computed the mean squared displacement (MSD) in the AA region of the H-AdResS simulation box. Only \ce{CO2} molecules in AA region are considered in this calculation, and since their number is not fixed and fluctuates with time, MSD is computed as a transient state measurement. As shown in Figure \ref{fig:msd}, the MSD curve computed from the H-AdResS simulation overlaps the reference fully atomistic simulations very well. In general, the self-diffusion coefficient of bulk \ce{CO2} gas is around $10^{-5} m^2/s$,\cite{Winkelmann2017}  whereas in ZIF-8 it is 5 orders of magnitude less than in the bulk ($10^{-9} - 10^{-10} m^2/s$\cite{Zhang2012,Krokidas2015}). Both reference and H-AdResS simulations yield a diffusion coefficient of $~1.5*10^{-9} m^2/s$\cite{Paudel2021}, in agreement with the previously reported measurements. This ensures once again that the dynamics in the AA region remains unaffected due to the coupling of different resolutions.

Finally, we measured the performance and scaling of our H-AdResS implementation. All the performance measurements are made on AMD EPYC Milan with 32 cores in the SACADO MeSU platform\cite{mesu}. For all the results discussed in the work, the size of AA region is 30\% of the total size of the box. For our main system of interest, ZIF-8/\ce{CO2} containing 9000 atoms, an average of 18 ns/day, standard deviation of 0.15 ns/day and peak performance of 19.06 ns/day were recorded in the reference AA simulations. In H-AdResS simulations, we recorded an average of 21 ns/day, standard deviation of $\sim$2 ns/day and peak performance of 23.87 ns/day, which is roughly 18-25\% higher than the reference AA simulations. As shown in previous works, the efficiency of the H-AdResS scheme depends on the ratio of AA region size in the total simulation box\cite{Potestio2013}.

To further develop this, we considered a system of fixed size, containing 30,000 atoms and simulated it in the H-AdResS scheme, by changing the size of the AA region. Considering a scheme with 15\% of its total size as AA region with an efficiency of 1, then the efficiency drops to 0.92 when the size of AA scheme is 35\%.  For every 5\% increase in the size of AA region, the efficiency dropped by 2\% (see Figure \ref{fig:efficiency}). This clearly shows the efficiency of H-AdResS simulations depends on the size of the AA region. This number is chosen as a trade-off between being able to measure the quantities of interest and profiting from the efficiency of the scheme. We additionally measured the scalability of our new implementation, by simulating a system of fixed size, with same settings but using an increasing number of processes and we found that the code scales linearly with number of processes (see Figure \ref{fig:scaling}).

\section{Conclusions}
In this work, we migrated the H-AdResS method to LAMMPS 2023, building on a previous implementation available for LAMMPS 2016. We implemented new pair styles to calculate forces and energies in the CG region. These new styles allow us to use different forms of force fields, from analytical forms such as simple Lennard Jones, to more sophisticated force fields developed using techniques such as IBI and FM, which are in tabular form and more generic CG force fields such as MARTINI. With these new pair styles, the method becomes more general and accessible to many users, removing the need for a specific type of CG force field. New i/o commands in the \textit{set} function enable us to assign values to the variables such as \textit{leader atom flag} and \textit{CG bead type}. With these features, we can directly manipulate these values from the LAMMPS input script to explore different CG mappings, without writing additional pre/post processing scripts. This makes the method more flexible, user friendly and offers less resistance for someone starting with the method for the very first time.

Since the position-based transition function is continuous in nature, simulating porous solids (that comprise regions that are void) within the H-AdResS scheme was an unsolved challenge. With the modifications that we introduced in the handling of the hybrid region in this implementation, the method can now be used for inhomogeneous density systems such as porous materials like MOFs and their gas adsorption. We validated our implementation by simulating bulk water, and we were able to reproduce results that are in very good agreement with our reference atomistic simulations as well as with results reported in the literature. With that confidence, we applied the method to a MOF, ZIF-8 both empty and loaded with \ce{CO2} gas. In both cases, the crystalline nature of the ZIF-8 is well reproduced and the structure remains intact. From the diffusivity and mean residence time measurements, it is clear that the dynamics in the AA region remains unaffected as well and can reproduce same quality results as a fully atomistic simulation. 

Finally, the performance, scaling and efficiency aspects of our implementation were discussed. The code scales linearly with number of processes, and the size of AA region influences the efficiency of H-AdResS simulations. Our new H-AdResS implementation yields $\sim$20\% increase in performance, compared to reference AA simulations. Future work will be devoted to extending the current implementation of H-AdResS to other interfaces, such as solid/liquid interfaces, relevant in the field of energy storage and electrocatalysis, as well as MOF/polymer interfaces for the membrane technologies field. This makes our new improved implementation of H-AdResS in LAMMPS an effective tool and will enable the efficient dual resolution simulations of heterogeneous systems with accurate structure and dynamic property estimations.

\section*{Acknowledgements}

The authors thank the Ecole Doctorale Chimie Physique et Chimie Analytique de Paris Centre for funding this work. R.S. thanks the European Research Council for an ERC StG (MAGNIFY project, number 101042514).  This work was granted access to the HPC resources of the SACADO MeSU platform at Sorbonne Université, where the simulations were performed. We thank Matej Praprotnik, Raffaello Potestio and Robin Cortes-Huerto for fruitful exchanges. We also thank Mathieu Salanne and Benjamin Rotenberg for useful discussions.

\section*{Data availability statement}

Data are made available in the Supporting Information. Input files, code and documentation will be made available upon publication in \url{https://github.com/hariharansudhakar/lammps-hadress}.

\section*{References}
\bibliography{iadress}

\clearpage
\newpage

\setcounter{equation}{0}
\setcounter{figure}{0}
\setcounter{table}{0}
\renewcommand{\theequation}{\roman{equation}}
\renewcommand{\thefigure}{S\arabic{figure}}
\renewcommand{\thetable}{S\arabic{table}}

\section*{Supplementary information}
\subsection{Simulation details} \label{sec:sim-details}
All the simulations were performed with a timestep of 1 fs. For liquid water, cutoff distances of 10 \AA \ for LJ interactions and 12 \AA \ for Coulomb interactions were used, whereas for ZIF-8 and ZIF-8/\ce{CO2} a cutoff distance of 13 \AA \ for both LJ and Coulomb interactions was used. Long-range interactions were calculated using the particle-particle particle-mesh (PPPM) method\cite{pppm}, with a relative error of $10^{-6}$. For all the reference AA simulations, Nosé-Hoover thermostat with damping constant of 0.1 ps and barostat with damping constant of 1 ps were used. All H-AdResS simulations were started without compensations. Both density and pressure compensation forces were updated every 1 ps. A damping constant $\alpha$ of 0.2 was used for the DSF potential. 
\subsection{H-AdResS parameters and compensation calculations}
In the main text, the choice of the size of the AA region ($l_{AA}$) and its effect of efficiency were discussed. Another important parameter is the size of hybrid region ($l_{HY}$). $l_{AA}$ is chosen in such a way that no atom in the AA region can directly interact with an atom in the CG region and vice versa. Therefore, a sensible choice for $l_{HY}$ could be, 1.5-2 times the intermolecular interaction cutoff distance. 

Pressure and density compensation routines are mostly implemented as in the 2016 implementation. Pressure compensation is computationally implemented as follows: first, the hybrid region is divided into small bins. The total number of bins depends on the choice of $\delta \lambda$, $N_{bins} = 1/\delta\lambda$ . For each atom $i$ of molecule $a$ present in a bin $b_i$, $U^{AA}$ and $U^{CG}$ are computed and accumulated in a local variable. Along with this, the total number of particles present in each bin is also recorded. The calculation is performed on all the molecules present in the hybrid region and the averages are updated at a regular time interval $\Delta t$. This calculated force is proportionately applied on each atom of a molecule, based on their mass fraction. 

To apply density compensation forces, the instantaneous density profile is computed at short regular intervals which leads to a noisy numerical gradient. To circumvent this problem, the simulation box is divided into small bins. The position of the center of mass of the molecule is smoothed out using a Gaussian function. This spreads the coordinates of the molecule across several bins instead of putting all of them into just one. In practice, this means that the density in a bin $i$, which covers the coordinate range [$x_i,x_{i+1}$] in a given simulation frame, is calculated based on a smoothed distribution\cite{Heidari2016}, following the equation (\ref{eq:gaussian-density}). Sigma ($\sigma$) determines how far the Gaussian is spread, it should be set roughly equal to the volume exclusion radius\cite{Praprotnik2008_2} of the molecule. $l$ is the cutoff and it is chosen in a way that it accounts for more than 98\% of the Gaussian surface. The rule of thumb is 2.5 to 3 times of sigma\cite{Allen2017}. Bin size is selected such that it is smooth enough and around 0.1 to 0.5 times the sigma value.\cite{Silverman2018}

 \begin{equation}\label{eq:gaussian-density}
 \begin{split}
\rho_i &= \sum_a \frac{1}{A} \int_{x_i}^{x_{i+1}} \exp\left[-\frac{(y - x_a)^2}{2\sigma^2}\right] dy \\
&\quad A = \int_{-l}^{l} \exp\left[ -\frac{y^2}{2\sigma^2} \right] dy
 \end{split}
\end{equation}

\subsection{Mean residence time calculation}
We define a spherical region centered in the centroid of the pore, which corresponds to the volume of the considered pore. The resulting pore radius is 5.8 \AA \ \cite{Weng2019}. When a given \ce{CO2} molecule stays inside the pore from $t=0$ to $t=t_i$, the variable H(t) will assigned 1 and 0 otherwise. \ce{CO2} molecules near the pore edge can vibrate across the geometric boundary that we defined, due to thermal fluctuations. Calculating residence time by simply taking the difference between the first exit time and the entry time would thus lead to underestimations. To accurately measure how long a gas molecule actually stays inside the pore, one must define a tolerance time,\cite{Impey1983} t*, that allows molecules to temporarily exit and re-enter without resetting their contribution to H(t).

\subsection{Radial distribution function calculations}

The structure was analysed by computing Radial Distribution Functions (RDFs) using custom python script. It was made to consider atoms present within spatially defined sub-region of the simulation box. The python script reads trajectories written by LAMMPS and flexible in selecting a specific time window and atom-type pairs. AA region of the simulation box was specified by Cartesian bounds along each axis. For each timestep, only atoms whose positions satisfy $x \in [x_{\min}, x_{\max}]$, $y \in [y_{\min}, y_{\max}]$, and $z \in [z_{\min}, z_{\max}]$ were considered for pair distance calculations. This spatially localized approach is particularly suited to heterogeneous systems such as confined fluids or electric double layers\cite{Nygrd2013,Horstmann2022,Zhang2024}, where global RDF estimates would obscure local structural features. To manage the computational cost of the \(\mathcal{O}(N^2)\) pairwise distance calculation across thousands of timesteps, the filtered frame set is partitioned into equal-sized chunks and distributed across all available CPU cores using Python's "multiprocessing.Pool". Each worker process independently accumulates a local pair-count histogram, which are subsequently aggregated into a single global count array upon completion. The resulting g(r) data, along with the raw histogram counts and bin centers, were written to plain-text files further analysis.

\begin{figure*}[h]
    \centering
    \includegraphics{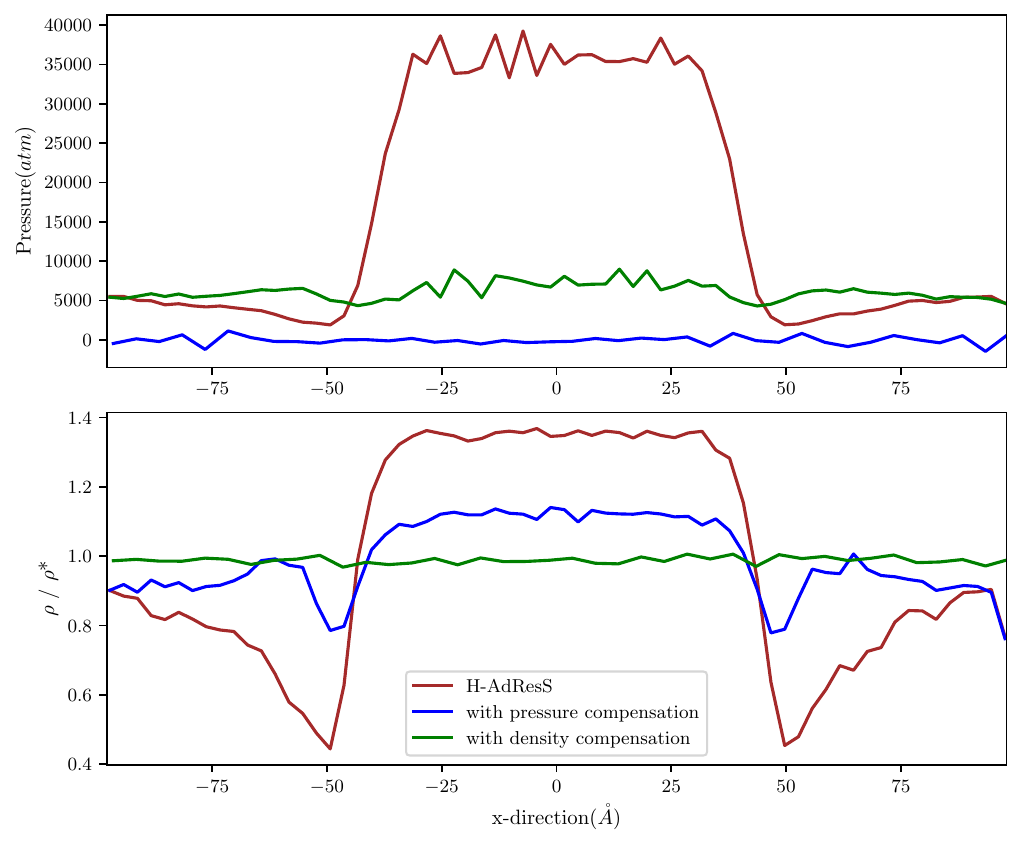}
    \caption{Pressure and density profiles of liquid water plotted from different H-AdResS setups. Simulations with no compensation activated result in a density imbalance and both pressure and density in the AA and CG regions (solid brown curves) are different. The blue (solid) curves represent a H-AdResS setup with pressure compensations active, and the green (solid) curves are of the setup with density compensations active.}
    \label{fig:dens-water}
\end{figure*}

 \begin{figure*}
    \centering
    \includegraphics[width=0.5\textwidth]{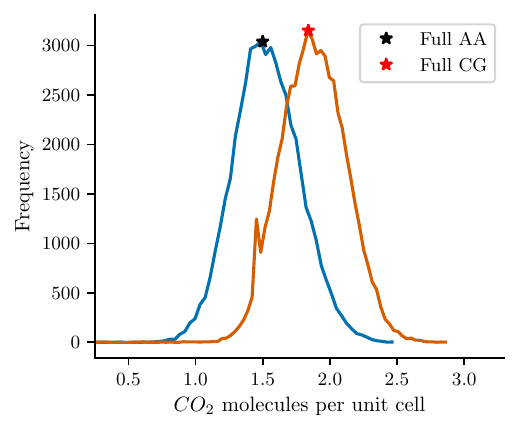}
    \caption{Histogram of \ce{CO2} molecule loading per unit cell of ZIF-8, measured from the GCMC+MD simulations. Orange and blue curves represent the CG and AA force fields respectively. Most favourable loadings reproduced by the CG (1.84 molecules/unit cell) and AA force fields (1.5 molecules/unit cell) are denoted by a red and a black star respectively.}
    \label{fig:loading}
\end{figure*}

 \begin{figure*}
    \centering
    \includegraphics[width=0.5\textwidth]{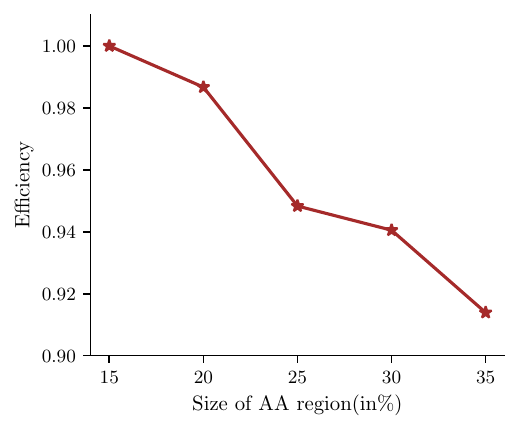}
    \caption{Efficiency as a function of the proportion of the AA region within the simulation box.}
    \label{fig:efficiency}
\end{figure*}

 \begin{figure*}
    \centering
    \includegraphics[width=0.5\textwidth]{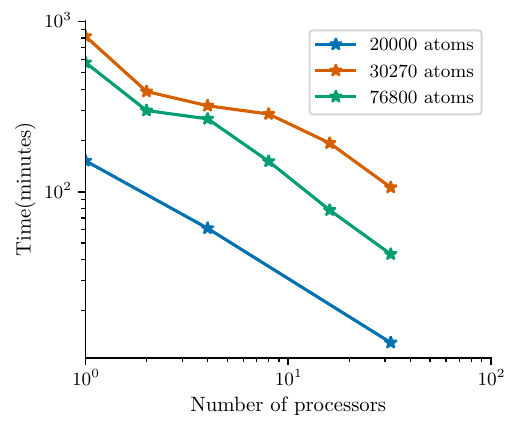}
    \caption{Scaling of our new H-AdResS implementation is measured by simulating a box of fixed size with an increasing number of processes, from 1 up to 32. Three different box sizes were considered.}
    \label{fig:scaling}
\end{figure*}

\end{document}